# Impact disruption of Bjurböle porous chondritic projectile

Tomas Kohout[1][2][3], Maurizio Pajola[4], Assi-Johanna Soini[5][6], Alice Lucchetti[7], Arto Luttinen[8], Alexia Duchêne[9], Naomi Murdoch[10], Robert Luther[11], Nancy L. Chabot[12], Sabina D. Raducan[13], Paul Sánchez[14], Olivier S. Barnouin[15], Andrew S. Rivkin[16]

Abstract

The ~200 m/s impact of a single 400-kg Bjurböle L/LL ordinary chondrite meteorite onto sea ice resulted in the catastrophic disruption of the projectile. This resulted in a significant fraction of decimeter-sized fragments that exhibit power law cumulative size and mass distributions. This size range is underrepresented in impact experiments and asteroid boulder studies. The Bjurböle projectile fragments share similarities in shape (sphericity, and roughness at small and large scale) with asteroid boulders. However, the mean aspect ratio (3D measurement) and apparent aspect ratio (2D measurement) of Bjurböle fragment is 0.83 and 0.77, respectively, indicating that Bjurböle fragments are more equidimensional compared to both fragments produced in smaller scale impact experiments and asteroid boulders. These differences may be attributed either to the fragment source (projectile vs. target), to the high porosity and low strength of Bjurböle, to the lower impact velocity compared with typical asteroid collision velocities, or potentially to fragment erosion during sea sediment penetration or cleaning.

[1] Corresponding author tomas.kohout@helsinki.fi
[2] Department of Geosciences and Geography, University of Helsinki, Finland
[3] Institute of Geology of the Czech Academy of Sciences
[4] INAF Astronomical Observatory of Padova, Padova, Italy
[5] Department of Geosciences and Geography, University of Helsinki, Finland
[6] Finnish Museum of Natural History, University of Helsinki, Finland
[7] INAF Astronomical Observatory of Padova, Padova, Italy
[8] Finnish Museum of Natural History, University of Helsinki, Finland
[9] Institut Supérieur de l'Aéronautique et de l'Espace (ISAE SUPAERO), Université de Toulouse, France
[10] Institut Supérieur de l'Aéronautique et de l'Espace (ISAE SUPAERO), Université de Toulouse, France
[11] Museum für Naturkunde, Berlin, Germany
[12] Johns Hopkins University Applied Physics Laboratory, Laurel, MD, USA
[13] Space Research and Planetary Sciences, Physikalisches Institut, University of Bern, Switzerland
[14] Colorado Center for Astrodynamics Research, University of Colorado Boulder, Boulder, USA
[15] Johns Hopkins University Applied Physics Laboratory, Laurel, MD, USA
[16] Johns Hopkins University Applied Physics Laboratory, Laurel, MD, USA

# 1. Introduction

Understanding of the impact disruption mechanism of planetary bodies and knowledge of fragment size and shape distributions are important to planetary defense mitigation efforts. They enable the prediction of fragment populations from natural planetary collisions, asteroid deflection or disruption efforts using kinetic impactors, as well as from impacts of small asteroids on the Earth. They also enable us to determine if an observed fragment population originates from a single impact event and to detect any subsequent fragmentation (e.g., thermal cracking) or sorting (e.g., gravitational transport or rotational ejection) processes altering the original fragment population.

The knowledge of impact disruption process comes from both observational and experimental studies with the caveat that there is a two order of magnitude gap in the spatial scale that these studies typically cover. The debris of both targets (Capaccioni et al. 1986; Capaccioni et al. 1984; Davis & Ryan 1990; Durda et al. 2015; Fujiwara, Kamimoto, & Tsukamoto 1977; Michikami et al. 2016; Michikami et al. 2018) and projectiles (Avdellidou et al. 2016; Daly & Schultz 2016; Kline & Hooper 2019; Wickham-Eade et al. 2018) studied during impact experiments are typically in μm-cm scale with either mass frequency distribution or size frequency distribution functions available while global observations of boulders and their size frequency distribution on asteroid surfaces cover sizes of $10^0$-$10^1$ m (DellaGiustina et al. 2019; Michikami & Hagermann 2021; Michikami et al. 2019; Michikami, Nakamura, & Hirata 2010) with smaller dm-sized boulders limited to few areas with high resolution imaging available (e.g., (Burke et al. 2021)). Because a significant fraction of small boulders is partly buried within regolith, their dimensions (especially their major axis) are not well constrained. In contrast the larger boulders may be positioned by seismic shaking to rest with their major *a* axis being parallel with the surface (Michikami & Hagermann 2021; Michikami, et al. 2019; Michikami, et al. 2010). While the two shorter *b* and *c* axes of big boulders can be often distinguished from 3D models, small boulder *b* and *c* axes are often not properly distinguished due to a lack of sufficient 3D imaging resolution with respect to their scale. Thus, the observed size frequency distribution function of the small size fraction often apparently deviates from the global value of $10^0$-$10^1$ m-sized boulders(Michikami, et al. 2019). This together with boulder thermal cracking, micrometeorite bombardment, migration, or ejection processes occurring on asteroids often creates bias in these studies.

In this study we aim to present missing information on impact disruption of chondritic material at dm scale in order to fill the existing gap in the experimental and observational data. In contrast to previously published work, we will investigate relatively weak and porous chondritic material and focus on catastrophic disruption of the projectile rather than the target. For this purpose, we present data on a unique historic impact disruption event of a single, 400-kg large, Bjurböle L/LL ordinary chondrite projectile producing a significant fraction of fragments in the dm range (Ramsay & H. 1902). As the mass, volume and 3D shape of fragments produced by this impact can be fully measured in the laboratory, the uncertainties in fragment dimensions and their orientations are reduced avoiding usual uncertainties in these parameters derived from remote sensing.

## 2. Methods

### 2.1. Bjurböle meteorite properties

Bjurböle meteorite is classified as L/LL4 ordinary chondrite of S1 shock stage (Catalogue of Meteorites (MetCat) 2022; Database 1994-2017; Database 2005-2023). Bjurböle is a friable meteorite of a high porosity (~20%, Table 1 and S1,(Flynn 2005; Pesonen, Terho, & Kukkonen 1993), above L/LL average of 5-9% (Consolmagno, Britt, & Macke 2008)) and of low longitudinal wave velocity $v_p$ (1030 m/s, (Flynn 2005)) pointing to relatively low Young's modulus and strength.

### 2.2. Fall circumstances

Bjurböle meteorite fall occurred in the evening hours of March 12, 1899 and was witnessed by many residents across the Baltic region. Ramsay and H. (1902) provides a detailed description of the fall circumstances of which we provide here brief summary. The trajectory roughly followed the southern Finnish coastline from west towards east. Several break-up events were observed along the luminous fireball trajectory. The terminal mass fell south of Finnish city of Porvoo (approx. 50 km east of Helsinki) on a frozen sea bay. A single 4-m sized hole was found in the ice surrounded by an 8-m circle of fractured ice and a 20-25 m rim of splashed water and sediment (Fig. 1). The fractured ice and ejecta rims are slightly extended towards the west, indicating an inclined impact from the eastern direction. This is consistent with a deceleration of the terminal mass towards free fall in the dark flight phase and reverse of its trajectory from westward to eastward by prevailing eastern low-level winds reported on the evening of the fall. The fall area consisted of a ~40 cm thick sea ice sheet over a ~1 m layer of water, followed by an ~8 m thick layer of unconsolidated muddy sea sediment.

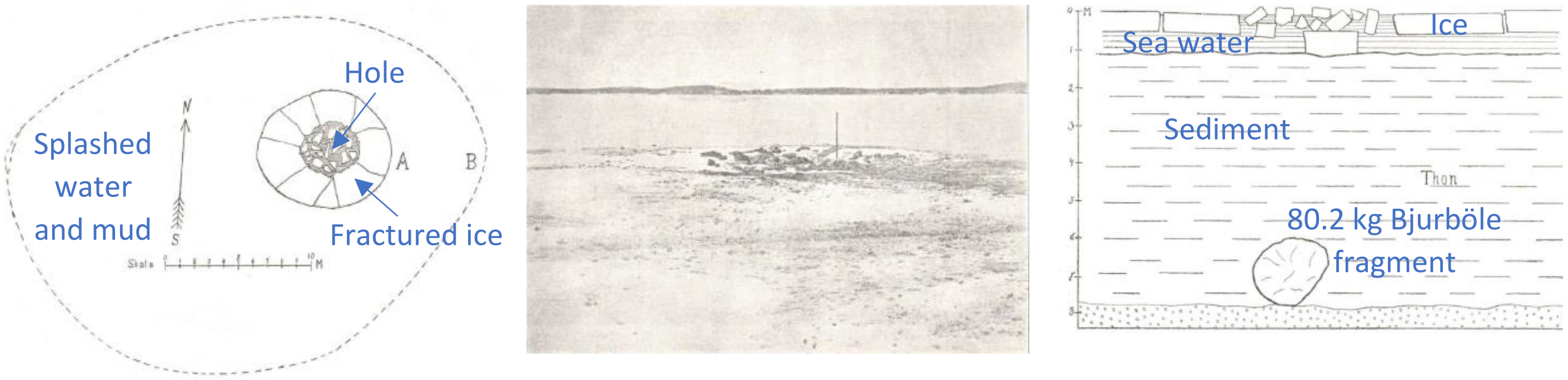

Fig. 1: Sketch of the fall area (left) photo (center), and a vertical section (right). From Ramsay and H. (1902), modified.

Recovery efforts were initialized with the aim to recover the terminal mass from the sea. A wooden casting was constructed around the impact hole and water with mud was partly pumped out and filtered. The largest fragment recovered from the impact site weighed 80.2 kg and was located within the bottom sediment approximately 6-7 m below the ice surface. Several other fragments in the range of 10-20 kg were also recovered along with countless smaller fragments. Original detailed records of fragment counts and sizes are missing, however, the total recovered mass is reported to have been 328 kg. Ramsay and H. (1902) lists the four largest recovered fragments as 80.2, 21, 18, and 17 kg. The first two correspond to our Helsinki A3747 (4) and Stockholm 990433 catalogued fragments and clearly appear in multiple historical photographs taken through the years and decades after the fall. However, any other historical or current information about the latter two fragments was not located nor their presence can be inferred from any historical photographs. Thus, we lean towards possibility that these fragments newer existed and their listing may be erroneous. Ramsay and H. (1902) also admits that many small (most likely cm-sized or smaller) fragments were missed during the recovery efforts and additional fragments may have been taken unreportedly by hired staff as souvenirs. Thus, the estimated total mass of the terminal meteorite prior to its impact with ice is possibly around ~400 kg. Based on available records, no meteorite fragments were observed on the sea ice in the vicinity of the impact nor searched along the fireball trajectory.

The above-mentioned historical evidence implies that all recovered meteorites were part of a single ~400 kg terminal body which decelerated towards free fall velocity and stayed intact upon its impact with the sea ice. The fragmentation occurred only upon initial contact with sea ice and thus, can be considered as single impact/collision event. This fragmentation mechanism leading to the origin of recovered meteorite fragments differentiates Bjurböle from other meteorite falls where the recovered meteorites result from one or multiple atmospheric impact events and form a strewn field. As mentioned above, atmospheric fragmentation of Bjurböle did occur, but the "atmospheric breakup" fragments most likely ended up on distant sea ice and were not searched for.

## 2.3. Impact velocity

The free fall impact velocity of Bjurböle terminal body can be estimated from equation (1) where the force of gravity on the object on the left equals drag force on the right:

$$mg = \frac{c_d \rho v^2 A}{2} \quad (1)$$

Where $m$ is object mass (400 kg), $g$ is the acceleration of gravity (9.81 m/s$^2$), $c_d$ is the drag coefficient (0.47 for a sphere, 1.05 for a cube), $\rho$ is the medium density (1.280 kg/m$^3$), $A$ is the object cross section (0.322 m$^2$ or 0.272 m$^2$ for a 400 kg, 2840 kg/m$^3$ dense (Table S1) sphere or cube, respectively).

Solving the equation (1) for $v$ results in:

$$v = \sqrt{\frac{2mg}{c_d A \rho}} \quad (2)$$

and returns ~200 m/s for a sphere and ~147 m/s for a cube.

## 2.4. Peak shock pressure

Peak shock pressure was estimated using the planar impact approximation (Melosh 2013) (Fig. 2). The Hugoniot pressures are determined for ice using the linear relation between shock wave speed $U$ and particle speed $u_P$:

$$\mathrm{U} = \mathrm{C} + \mathrm{S} * \mathrm{u}_P \quad (3)$$

with $C$ being 1317 m/s and $S$ 1.526 at -15°C for target (ice) (Melosh 2013). For the Bjurböle meteorite, the linear relation is not known, but $C$ can be assumed as value of the bulk speed of sound 1030 m/s (Flynn 2005). $S$ was assumed as 1.42, similar to sandstone, which has similar values of porosity and speed of sound, but also larger values of *S* like for basalt do not significantly affect the result.

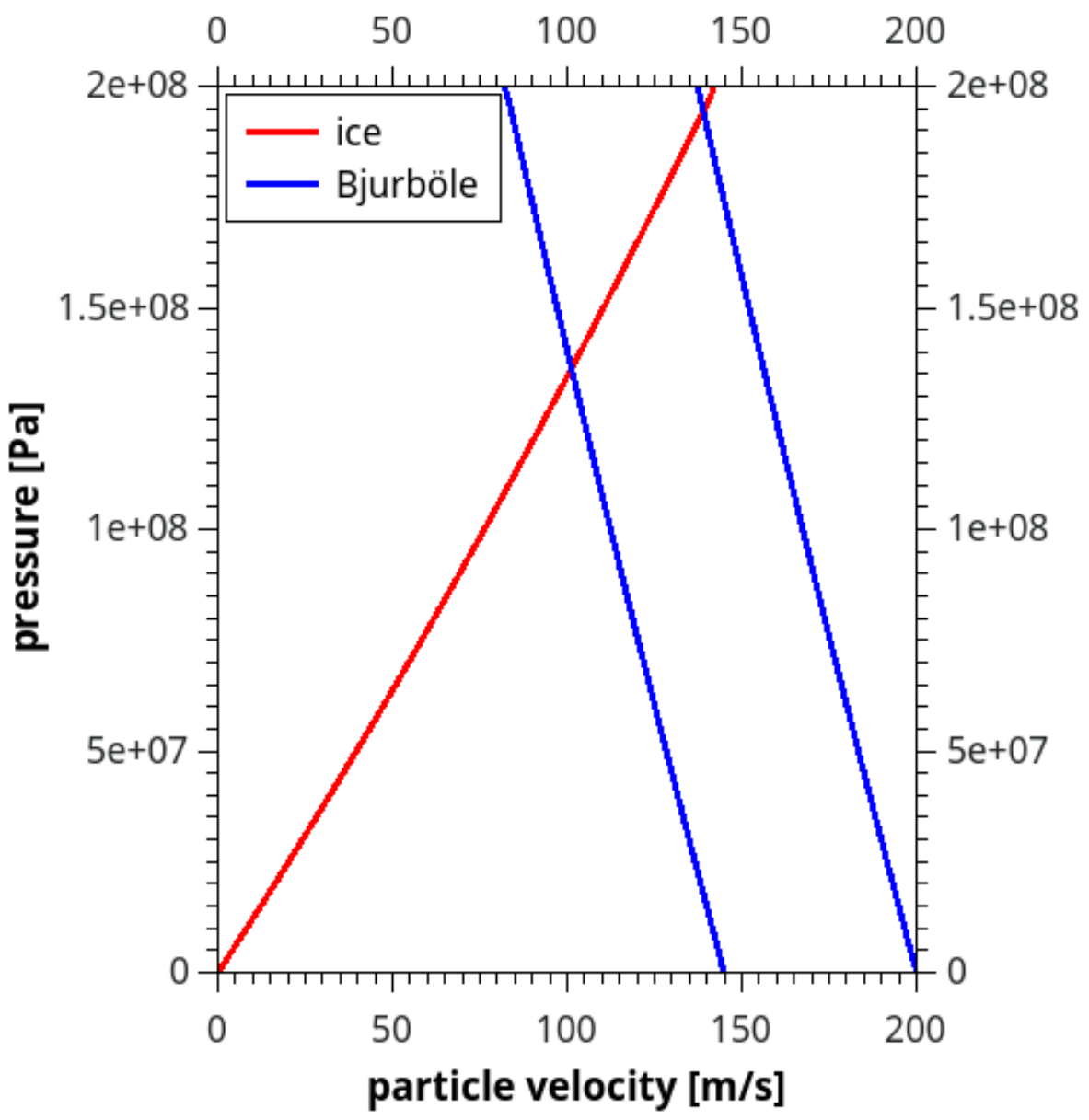

Fig. 2.: Peak shock pressure estimation using the planar impact approximation. The red curve shows the Hugoniot curve for ice. The blue curves show the Hugoniot curve estimates for the Bjurböle meteorite and are plotted for impact speeds of 147 m/s (cube) and 201 m/s (sphere).

## 2.5. Fragment characterization

For the size and shape distribution study we gathered information on the size and mass of nearly all Bjurböle meteorites larger than 1 kg reported to be present in collections worldwide. For this we used the Finnish Natural History Museum (LUOMUS) historical records as well as the meteorite list in MetBase database with 28 samples reported. The meteorite list is presented in Table 1. The size and mass of meteorites in the LUOMUS collection as well as of the 20-kg sample from Swedish Museum of Natural History (20 in total) were determined from 3D models and using calibrated scales (mass reported with valid digits of the calibration), respectively. For 3D model measurements, a NextEngine 3D Scanner Ultra HD model 2020i laser scanner and for shape reconstruction a NextEngine ScanStudio software was used. The volume and bounding box dimensions were measured in Geomagic Verify Viewer software. Dimensions of bounding box were used as fragment major *a*, minor *c*, and intermediate *b* perpendicular axes ($a \geq b \geq c$). Axes dimension and mass of the other samples dispersed around the world were measured manually and provided upon request by the museum curators or owners.

Table 1. The physical properties of the studied Bjurböle meteorites together with information on fusion crust presence. The bulk volume, bulk density, and porosity are only available for the laser-scanned samples.

| Meteorite | Mass (g) | Bulk Volume (cm³) | Bulk Density (g/cm³) | Porosity (%) | a (mm) | b (mm) | c (mm) | b/a | c/a | c/b | Eq.d. (mm) | Fusion crust |
|---|---|---|---|---|---|---|---|---|---|---|---|---|
| Helsinki A3747 (4) | 80200 | 29172 | 2.75 | 22.1 | 518 | 385 | 335 | 0.74 | 0.65 | 0.87 | 378 | 1 face with irregular patches |
| Stockholm 990433 | 20900 | 7302 | 2.86 | 18.9 | 344 | 290 | 199 | 0.84 | 0.58 | 0.69 | 241 | 2 faces with irregular patches |
| Helsinki D6147 (AL79) | 13480 | 4861 | 2.77 | 21.4 | 260 | 223 | 212 | 0.86 | 0.82 | 0.95 | 209 | |
| Helsinki combined D6164 + D6152 # | 10643 | | | | 238 | 195 | 133 | 0.82 | 0.56 | 0.68 | 193 | 2 faces |
| Helsinki D6149 (AL86 / 33 5413) | 5450 | 1877 | 2.90 | 17.8 | 179 | 167 | 148 | 0.93 | 0.83 | 0.89 | 154 | 7-cm patch |
| Helsinki D6150 (TK1) | 5300 | 1863 | 2.85 | 19.4 | 203 | 186 | 116 | 0.92 | 0.57 | 0.62 | 153 | 10x3-cm patch |
| Helsinki D6148 (AL81) | 4350 | 1518 | 2.86 | 18.8 | 200 | 170 | 131 | 0.85 | 0.66 | 0.77 | 143 | 5-cm patch on one face |
| Kettunen-Moilanen RH80 (Haag 107) # | 3300 | | | | 160 | 142 | 116 | 0.89 | 0.73 | 0.82 | 130 | |
| Helsinki D6163 (3130 / 26) | 3113.7 | 1083 | 2.88 | 18.5 | 163 | 132 | 118 | 0.81 | 0.72 | 0.89 | 128 | |
| Chicago ME 1428 #4 | 2995.5 | | | | 155 | 120 | 110 | 0.77 | 0.71 | 0.92 | 126 | 1 face |
| St Petersburg Univ. Geol. 4371 | 2693 | | | | 160 | 120 | 100 | 0.75 | 0.63 | 0.83 | 122 | 3-cm patch |
| Oslo MET 5 | 2095 | | | | 160 | 115 | 85 | 0.72 | 0.53 | 0.74 | 112 | 1 face + 4-cm patch |
| Vienna NHMW-H1462 | 1939.4 | | | | 140 | 115 | 80 | 0.82 | 0.57 | 0.70 | 109 | |
| Helsinki D6151 (AL2 27) | 1879 | 665 | 2.82 | 20.0 | 149 | 116 | 87 | 0.78 | 0.58 | 0.75 | 108 | |
| Helsinki D6146 (AL68) | 1809.4 | 639 | 2.83 | 19.8 | 153 | 133 | 88 | 0.87 | 0.58 | 0.66 | 107 | |
| Killgore CH-1-C | 1788 | | | | 166 | 130 | 62 | 0.78 | 0.37 | 0.48 | 106 | 2-cm patch |
| Helsinki D6144 (AL32 / 143) | 1704.8 | 590 | 2.89 | 18.1 | 161 | 97 | 90 | 0.60 | 0.56 | 0.93 | 105 | |
| Chicago ME 1427 #1 | 1611.82 | | | | 140 | 110 | 80 | 0.79 | 0.57 | 0.73 | 103 | |
| Oslo MET 6 | 1570 | | | | 135 | 120 | 110 | 0.89 | 0.81 | 0.92 | 102 | 1 face |
| Senckenberg MPI 235/10 | 1495 | | | | 135 | 105 | 80 | 0.78 | 0.59 | 0.76 | 100 | |
| St Petersburg mining museum 92-1 | 1441 | | | | 140 | 100 | 80 | 0.71 | 0.57 | 0.80 | 99 | 3-cm patch |
| Porvoo 9184 | 1432.8 | 504 | 2.84 | 19.4 | 126 | 124 | 86 | 0.98 | 0.68 | 0.69 | 99 | 2 2-cm patches |
| Aalto 4606 | 1364.7 | 469 | 2.91 | 17.5 | 126 | 98 | 92 | 0.78 | 0.73 | 0.94 | 97 | 7x5-cm patch |
| Rose D5504 | 1365 | | | | 110 | 90 | 80 | 0.82 | 0.73 | 0.89 | 97 | |
| Helsinki D6145 (AL67 / 152) | 1358.6 | 474 | 2.86 | 18.9 | 124 | 117 | 70 | 0.94 | 0.56 | 0.60 | 97 | |
| Cluj 1.308 | 1300 | | | | 120 | 110 | 80 | 0.92 | 0.67 | 0.73 | 96 | 1 face |
| Helsinki D6143 (AL31 / 144) | 1237.9 | 442 | 2.80 | 20.7 | 109 | 104 | 90 | 0.95 | 0.83 | 0.87 | 94 | |
| Stockholm 540022 | 1176.7 | 405 | 2.91 | 17.7 | 94 | 93 | 93 | 0.99 | 0.99 | 1.00 | 92 | |
| Oxford OUMNH-Mt006 | 1168 | | | | 104 | 89 | 92 | 0.86 | 0.88 | 1.03 | 92 | 10x2-cm patch |
| Bartoschewitz BC8.2 | 1157 | | | | 130 | 90 | 70 | 0.69 | 0.54 | 0.78 | 92 | |
| Vernadsky 87 | 1135.5 | | | | 110 | 100 | 80 | 0.91 | 0.73 | 0.80 | 91 | 1 face with irregular patch |
| Cureton 1113 * | 1113 | | | | | | | | | | 91 | |
| Vernadsky 15155 | 1090 | | | | 110 | 100 | 75 | 0.91 | 0.68 | 0.75 | 90 | |
| Porvoo 88-38 | 1069.6 | 377 | 2.83 | 19.7 | 117 | 96 | 62 | 0.82 | 0.53 | 0.65 | 90 | 9x1-cm and 2x1-cm patches |
| Smithsonian USNM695 | 1045 | | | | 75 | 72 | 69 | 0.96 | 0.92 | 0.96 | 89 | |
| Paris 2259don | 1043.2 | | | | 100 | 90 | 70 | 0.90 | 0.70 | 0.78 | 89 | |
| Mean | | | | | | | | 0.84 | 0.67 | 0.80 | | |
| s.d. | | | | | | | | 0.09 | 0.13 | 0.12 | | |
| Helsinki D6164 (AL80 / 5318) | 5150 | 1819 | 2.83 | 19.8 | 238 | 195 | 125 | | | | | 2 faces |
| Helsinki D6152 (AL80) | 4500 | 1586 | 2.84 | 19.6 | 220 | 187 | 133 | | | | | 2 faces |
| Kettuen RH80 (Haag 107) without corner | 2977.6 | 1010 | 2.95 | 16.5 | 148 | 142 | 116 | | | | | |
| Moilanen RH80 (Haag 107) detached corner | N/A | N/A | | | | | | | | | | |
| Mean | | | 2.85 | 19.2 | | | | | | | | |

| | | | | |
|---|---|---|---|---|
| s.d. | | | 0.05 | 1.4 |

Notes:

* - Current location and information on dimensions is unknown.

# - Sample split. Individual fragments are listed at the end of the table.

The largest 80.2 kg Bjurböle sample is currently partly embedded in a plaster support (lower ~5 cm). Thus its vertical dimension and mass were not possible to determine. However, a high-fidelity plaster model of the meteorite exists. We scanned both the real meteorite and its plaster model and verified that both horizontal dimensions agree within 1.8 (*a*) and 6.6 (*b*) cm resulting in estimated measure error of ~ 3% (*a*) and 17% (*b*). Horizontal dimensions were derived from the real meteorite while the vertical (*c*) dimension and volume were derived from the plaster 3D model. As mass of the sample, we used the 80.2 kg value found in the LUOMUS catalogue.

In some of the experiments or the asteroid observations reported in the literature, the fragment dimensions are measured from 2D images by fitting a bounding box or ellipse where it is difficult to distinguish the intermediate *b* and the minor *c* axis from each other as well as to estimate the distortion of major *a* axes due to fragment tilt. In this case we refer to these values in the discussion as to apparent *a′* and *b′* dimensions and *b′*/*a′* ratio as to apparent aspect ratio. To simulate a determination of the bounding box axes in 2D with Bjurböle samples, we determined the apparent aspect ratio (apparent *b′*/*a′* ratio) as average bounding box width / length of three random 2D orthogonal projections of the 3D Bjurböle shape models.

To mimic asteroid boulder observations the random 2D orthogonal projections of the 3D Bjurböle shape models were also used to analyze several adimensional shape and morphological parameters such as the roundness, circle ratio sphericity, circularity, solidity, and the equivalent diameter using the 2D shape processing pipeline described in Robin (2024). These morphological descriptors are defined in Table 2. The circle ratio sphericity is the radius ratio of the maximum inscribed and the minimum circumscribing circle. The roundness measures the curvature of corners in a particle; a value of 0 and 1 indicates a very angular and rounded particle, respectively. The circularity and solidity indicate the large- and small-scale roughness, respectively. The circularity is a perimeter ratio, and the solidity refers to the convexity of the particle. Using the morphology analysis derived from the 2D projections of the shape models enables removing the bias of the 3D representation compared to other results obtained from 2D morphology analysis.

Some of these parameters are dependent on the resolution (e.g., the particle is resolved by insufficient amount of pixels), as for the roundness (Zheng & D. 2015) and circularity(Nagaashi, Aoki, & Nakamura 2021). Both the 3D representation and its 2D projections have been acquired at high resolution (0.1-0.01 mm/px for the 2D projections) and enable the morphological analysis to be free of any biases resulting from the low resolution effect.

Table 2: Morphological descriptors computed for the 2D projections analyses with their equation, definition and reference (if applicable).

| Morphological descriptor | Equation | Variables definition | Reference |
|---|---|---|---|
| Circle ratio sphericity | $\frac{R_{ins}}{R_{cir}}$ | $R_{ins}$, the radius of the maximum inscribed circle; $R_{cir}$, the radius of the minimum circumscribed circle. | (Riley 1941) |
| Roundness | $\frac{\sum_{i=0}^{N} r_i / N}{R_{ins}}$ | $N$, the number of corners; $r_i$, the radius associated with each corner. | (Wadell 1932) |
| Circularity | $\frac{4\pi A}{P^2}$ | $A$, the projected area; $P$ , the perimeter of the particle. | (Cox 1927) |
| Solidity | $\frac{A}{H}$ | $H$, the area of the convex hull. | (Mora & Kwan 2000) |
| Apparent aspect ratio | $\frac{b'}{a'}$ | $b'$, the width (smaller axes of the 2D bounding box); $a'$, the length (longest axes of the 2D bounding box). | - |
| Equivalent diameter | $2\sqrt{\frac{A}{\pi}}$ | $A$, the area of the projected shape. | - |

The volume and mass of the meteorites were used to calculate their bulk density. We report new bulk density measurements of 27 Bjurböle fragment ranging in mass from 12.25 g up to 80.2 kg (Table 1 with meteorites larger than 1 kg, supplementary Table S1 with all meteorites and density statistics). Four smaller Bjurböle samples (Helsinki small 1-4) ranging from 17.27 g up to 21.62 g were used also for grain density measurement using an Quantachrome Ultrapyc 1200e gas pycnometer and $N_2$ gas. The average grain density 3.53 g/cm$^3$ of these small samples agrees within s.d. with the mean grain density of L and LL chondrites (3.51±0.11 g/cm$^3$ and 3.48±0.08 g/cm$^3$, respectively) and was used to calculate porosity of the larger Bjurböle samples which did not fit the gas pycnometer chamber.

For comparison, additionally we used average bulk density 3.29 g/cm$^3$ and grain density 3.47 g/cm$^3$ of 40 smaller Chelyabinsk samples (average of all lithologies combined) ranging in mass from 4.35 g up to 310.60 g with average mass of 54.40 g reported in (Kohout et al. 2014) together with bulk volumes and masses of three additional larger samples presented in (Kohout et al. 2017) (Table S1).

## 2.6. Power law fit

A power law fit of the major *a* axis and mass values was done following Clauset, Shalizi, and Newman (2009) technique which validates the existence of the power law. The method returns a scaling parameter $\alpha$ (power-law index) and a completeness limit $x_{min}$. The estimation of $x_{min}$ is done through the Kolmogorov-Smirnoff (KS) statistic and allows to find the $x_{min}$ value minimizing it. Afterwards, the parameter is determined through the maximum likelihood estimator. The uncertainty for both $\alpha$ and $x_{min}$ is then derived through a non-parametric bootstrap procedure that generates a large number of synthetic datasets from a power-law random generator and performs a number of KS tests to verify if the generated and observed data come from the same distribution. This technique returns a *p*-value that can be used to quantify the plausibility of the hypothesis. Considering the significance level of 0.10, if the *p*-value is >0.1, then it is possible to conclude that any difference between the empirical data and the model can be explained with statistical fluctuations. On the contrary, if the *p*-value is <0.1, then the data set does not come from a power-law distribution, but instead, from a different one.

# 3. Results

## 3.1. Impact circumstances

Free fall calculation for the ~400 kg terminal Bjurböle body estimates impact velocity at the end of dark flight in approximate range of 145-200 m/s for cubic and spherical shapes. The associated peak shock pressure reached within the Bjurböle projectile is ~ 140-200 MPa. The ratio of the largest surviving fragment (80.2 kg) to the projectile mass (~400 kg) is ~0.2. There is no indication of significant bulk density and porosity variations with fragment size from ~ 10 g up to the largest 80.2 kg indicating high homogeneity within the Bjurböle material at wide size range. Such high homogeneity in physical properties is similar to what is observed in Chelyabinsk LL fall known for co-existence of various shock lithologies, where a comparably wide size-range of fragments was studied (Kohout, et al. 2017) (Fig. 3).

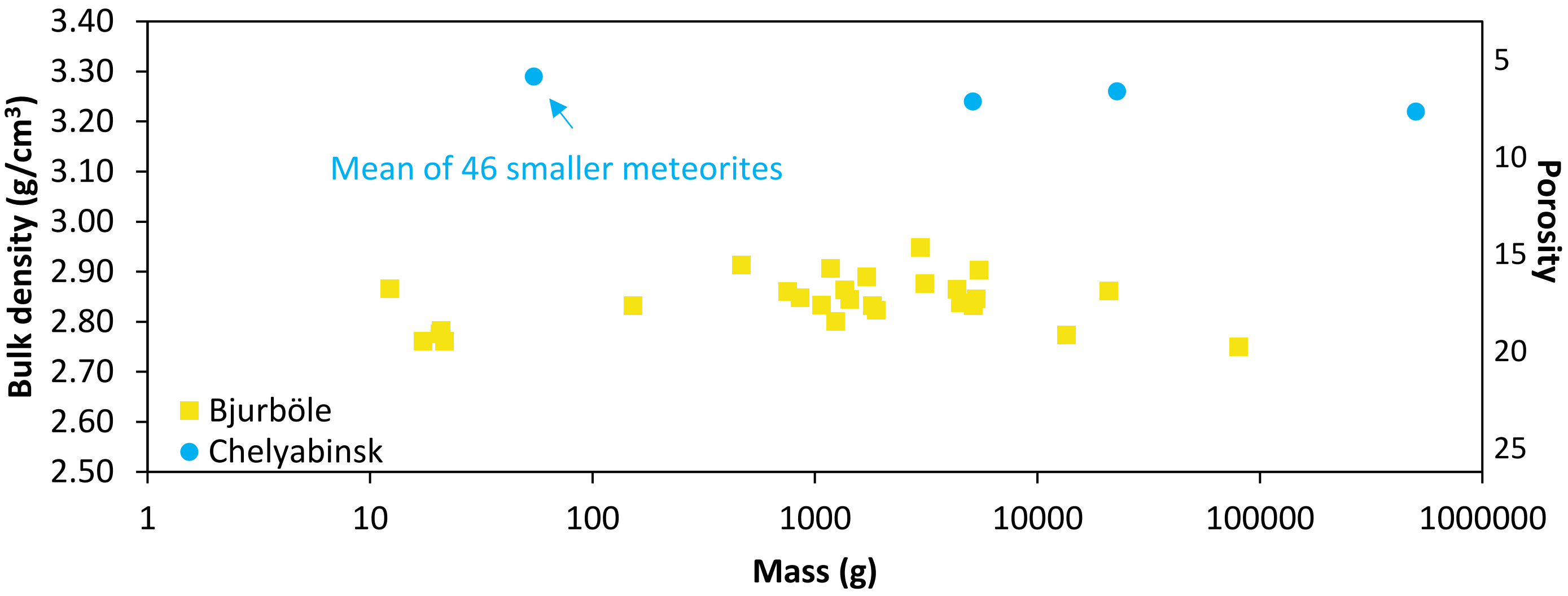

Fig. 3. Bulk densities and porosities of Bjurböle samples as a function of mass. Chelyabinsk data (Kohout, et al. 2014; Kohout, et al. 2017) are plotted for comparison.

### 3.2. Fragment dimensions and shapes

The results on fragment dimensions, mass, and bulk and grain density are summarized in Table 1. We managed to confirm mass of 36 and dimensions of 35 (to us) known Bjurböle fragments larger than 1 kg (totaling 188 kg), out of which 16 were laser-scanned and 19 were measured by hand with a measurement tape. Additionally we laser-scanned three fragments listed at the end of Table 1 which were accidentally broken during curation as detailed below.

In one case it was inferred from LUOMUS historic records that the fragments Helsinki D6164 and D6152 are broken pieces of a larger fragment with catalogue mass of 10.643 kg. In the dimension ratio and frequency mass / size distribution analyses we list this sample only once with each dimension represented by the estimated combined value and combined catalogue mass (being approx. 1 kg larger than the current mass of surviving fragments, most likely due to additional mass loss). For bulk and density and porosity statistics we use individual fragment masses and volumes.

In another case, the sample RH80 (Haag 107) (Robert Haag being previous owner) with a recent corner split and Jarkko Kettunen and Jarmo Moilanen being current owners of the two parts. Thus, we similarly used original combined dimensions and mass prior the split for dimension ratio and frequency mass / size distribution analyses while the new laser scan of the current larger part Kettunen RH80 is used for density and porosity statistics.

The Porvoo 9184 fragment does have also a detached corner. We were able to perform its laser scan with the corner in its original position obtaining representative results for its original shape.

Besides these three documented cases and occasional observed detachment of corners we did not find record of significant fragment split. We cannot rule out such historical occurrences. However, if these occur it will happen along preexisting loose crack being on the verge of natural fragmentation anyway. We do not assume this occurred on large scale in past and possible mass loss in order of few percent do not significantly affected our results.

The sample Cureton 1113 is listed in LUOMUS historical catalogue to be transferred to Curetons collections. Unfortunately, we did not find any information about current presence of the sample and thus, we cannot measure its dimensions. We, however, use the listed mass values in the fragment cumulative size distribution statistics.

We did extensive search for additional fragments private collections including request for information directed to renowned meteorite collectors and public announcement in the fall area through press release and local social media groups. Despite these efforts we were not able to locate additional fragments.

The dimmensions *b/a*, *c/a* and *c/b* ratios together with apparent aspect ratio distribution histograms are presented in Fig. 4. The mean *b/a* ratio is 0.84 with a s.d. (standard deviation) of 0.09, the mean *c/a* ratio is 0.67 with a s.d. of 0.13, and the mean *c/b* ratio is 0.80 with a s.d. of 0.12. All dimension ratios show unimodal distribution and the range of the apparent aspect ratio values is narrow centered towards left (lower values). The other shape and morphological parameters such as apparent aspect ratio, circularity, equivalent diameter, roundness, circle ratio sphericity, and solidity determined from 2D surface projections of the 15 laser-scanned Bjurböle fragments are listed in Table 3.

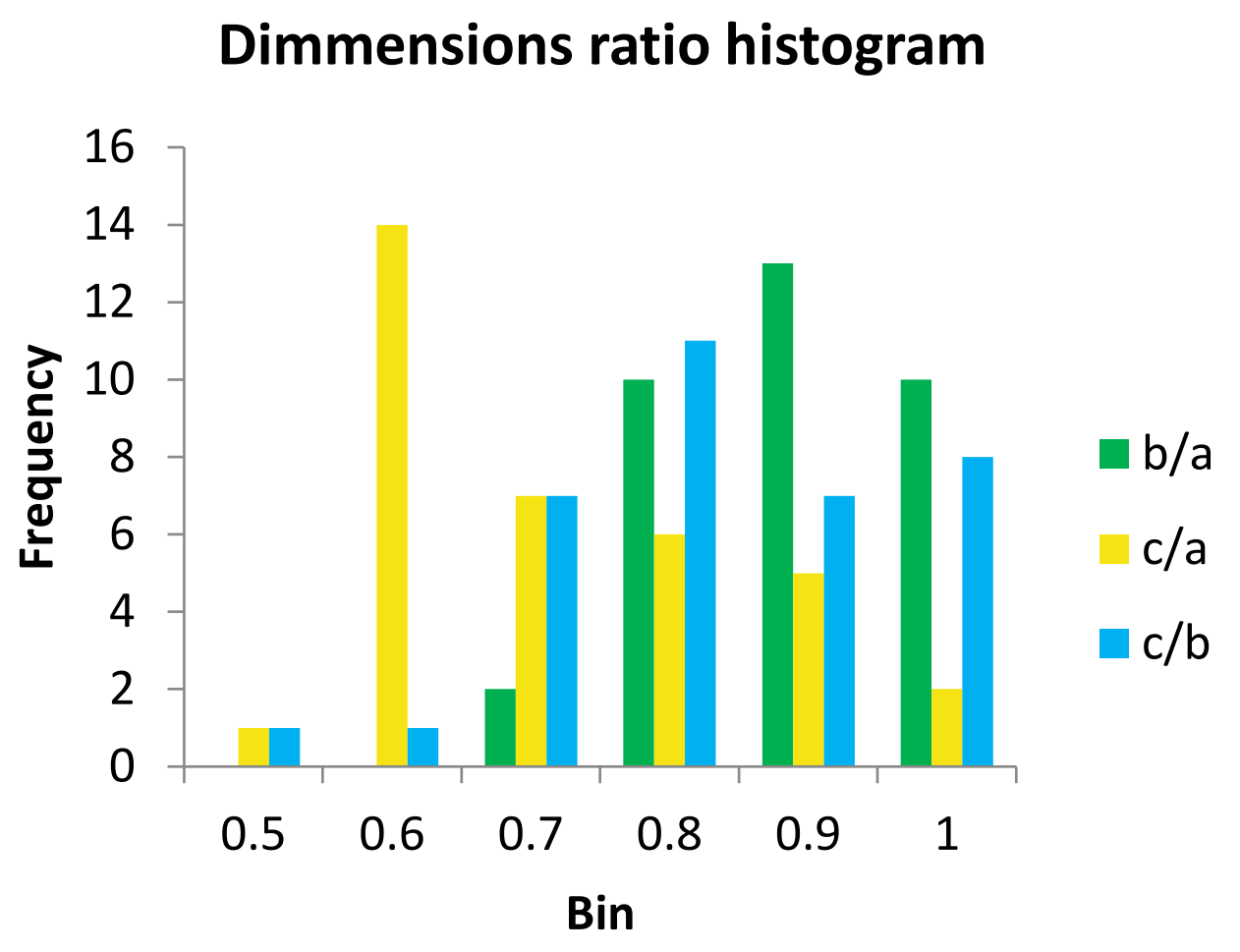

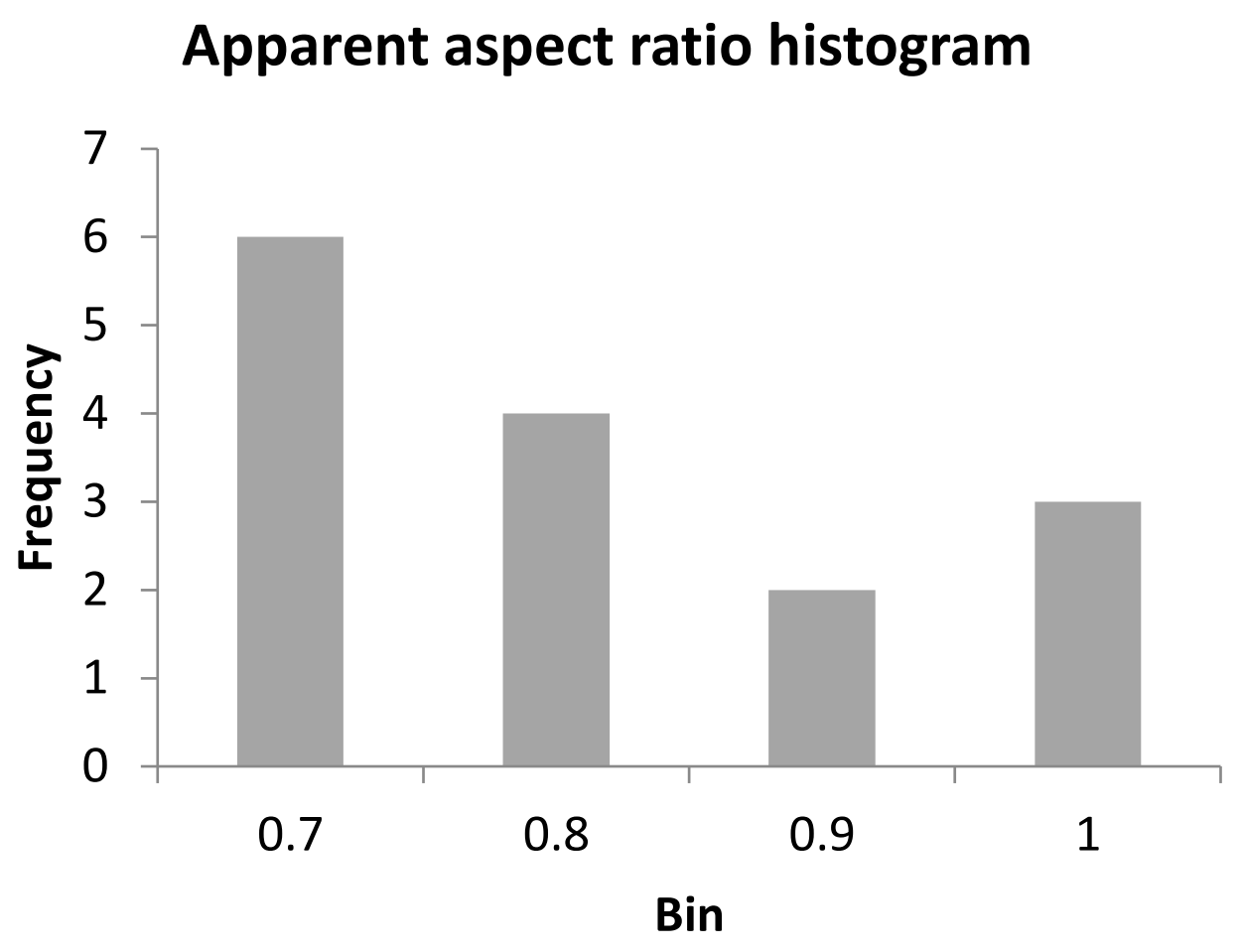

Fig. 4. Dimmensions *b/a*, *c/a*, and *c/b* ratios (left) and apparent aspect ratio (right) distribution histograms of Bjurböle fragments larger than 1 kg.

Table 3. Shape and morphological parameters determined from 2D surface projections of the 15 laser-scanned Bjurböle meteorites together with asteroid range from Robin (2024).

| Meteorite | Equivalent diameter (mm) | Apparent aspect ratio | Roundness | Sphericity | Circularity | Solidity |
|---|---|---|---|---|---|---|
| Helsinki D6143 (AL31 / 144) | 100 | 0.92 | 0.34 | 0.73 | 0.85 | 0.97 |
| Helsinki D6144 (AL32 / 143) | 108 | 0.69 | 0.31 | 0.60 | 0.77 | 0.96 |
| Helsinki D6145 (AL67 / 152) | 101 | 0.68 | 0.36 | 0.60 | 0.80 | 0.98 |
| Helsinki D6149 (AL86 / 33 5413) | 161 | 0.78 | 0.36 | 0.70 | 0.87 | 0.98 |
| Stockholm 540022 | 95 | 0.99 | 0.37 | 0.82 | 0.86 | 0.98 |
| Stockholm 990433 | 260 | 0.80 | 0.30 | 0.63 | 0.80 | 0.97 |
| Helsinki D6151 (AL2 27) | 113 | 0.69 | 0.29 | 0.62 | 0.79 | 0.97 |
| Helsinki D6163 (3130 26) | 134 | 0.84 | 0.35 | 0.72 | 0.84 | 0.98 |
| Porvoo 88-38 | 88 | 0.66 | 0.29 | 0.66 | 0.80 | 0.98 |
| Porvoo 9184 | 103 | 0.71 | 0.24 | 0.59 | 0.80 | 0.96 |
| Helsinki A3747 (4) | 407 | 0.69 | 0.33 | 0.63 | 0.82 | 0.97 |
| Helsinki D6146 (AL68) | 116 | 0.65 | 0.27 | 0.58 | 0.78 | 0.96 |
| Helsinki D6147 (AL79) | 226 | 0.90 | 0.32 | 0.76 | 0.83 | 0.97 |
| Helsinki D6148 (AL81) | 154 | 0.87 | 0.38 | 0.66 | 0.83 | 0.97 |
| Helsinki D6150 (TK1) | 161 | 0.70 | 0.36 | 0.64 | 0.84 | 0.98 |
| Mean | 155 | 0.77 | 0.32 | 0.66 | 0.82 | 0.97 |
| s.d. | 86 | 0.11 | 0.04 | 0.07 | 0.03 | 0.01 |
| Asteroid boulder range | | 0.70 - 0.75 | < 0.56 | 0.61 - 0.67 | 0.84 - 0.90 | 0.95 - 0.96 |

### 3.3. Fragment cumulative size and mass distribution

For consistency with earlier asteroid boulder studies or impact disruption experiments, we choose the meteorites' major *a* dimension and mass, respectively, to plot these against their cumulative number in a log-log space (Fig. 5) and use the Clauset, et al. (2009) technique to validate applicability of the power-law fit. In the case of major *a* dimension, the resulting "size" scaling parameter $\alpha$ (also called power-law index) is −2.7 ± 1.4 and the completeness limit $x_{min}$ is 130 ± 23 mm. In the case of mass, the "mass" $\alpha$ is −1.0 ± 0.4 and the completeness limit $x_{min}$ is 1135 ± 435 g. Considering the significance level of 0.1, the p-values we obtained computed from 5000 Kolmogorov-Smirnoff statistical tests are 0.78 and 0.75, respectively. This means that the dataset comes from a power-law distribution and any difference between the empirical data and the model can be explained solely with statistical fluctuations. Fig. 5 also depicts the frequency plot of the two datasets, together with the mode, median, mean, and other parameters of the distribution.

The $x_{min}$ of both datasets corresponds to cumulative count of ~22-24 meaning that in terms of dimension as well as of mass, ~80% of the meteorites fit the power law distribution.

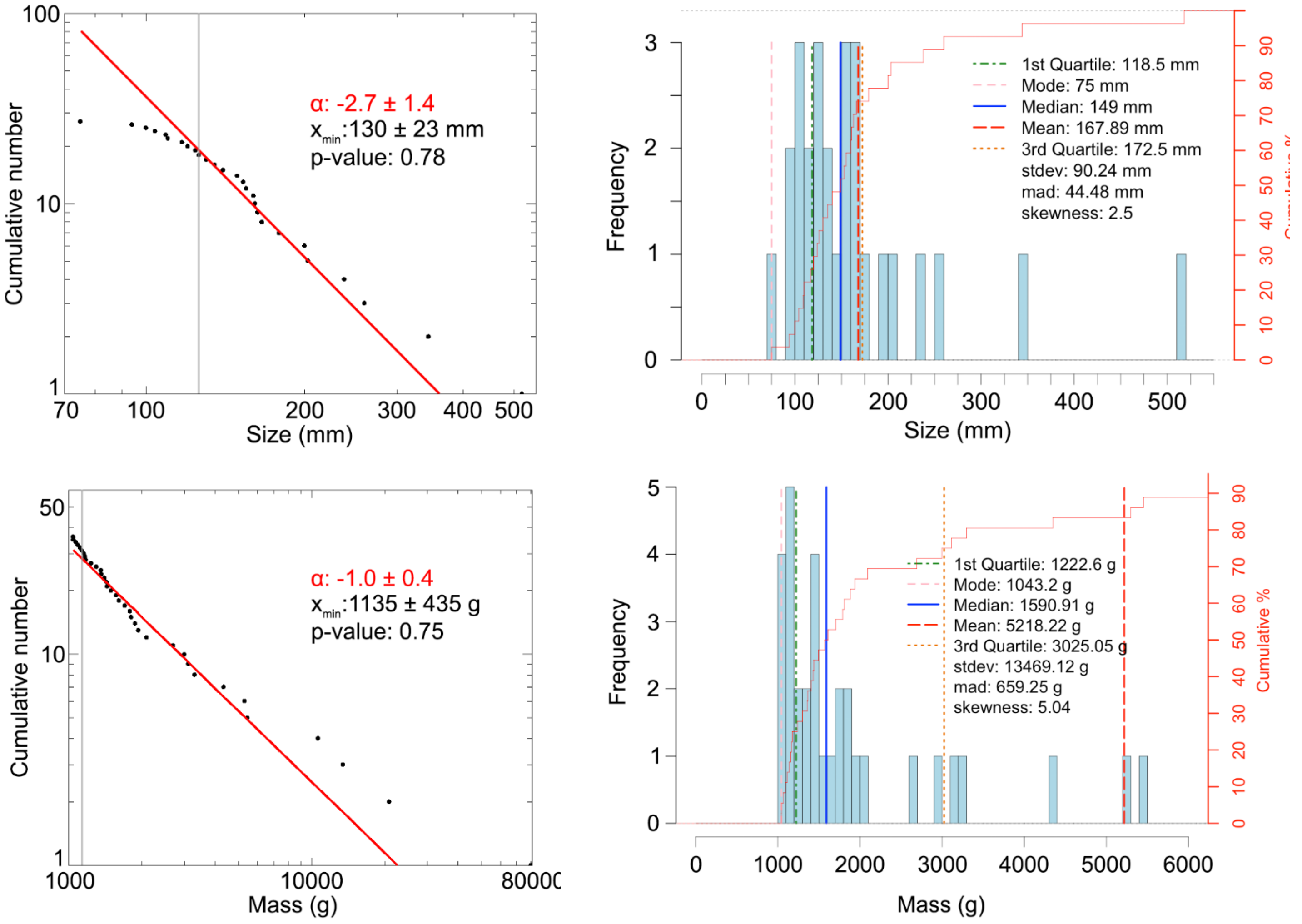

Fig. 5. Power law fit (left) and frequency and cumulative plots (right) of the major *a* dimension (up) and mass (down) distribution of the Bjurböle meteorites larger than 1 kg, together with best fit parameters.

## 4. Discussion

Bulk density measurements of all Bjurböle meteorites (supplementary Table S1) reveal that they are homogeneous from cm (10 g) up to dm (80 kg) size range (Fig. 3) with mean value of 2.84 g/cm$^3$ (s.d. 0.05) being lower than the mean L (3.36 g/cm$^3$) and LL (3.54 g/cm$^3$) values (Consolmagno, et al. 2008) due to Bjurböle's above-average porosity of 19.6% (s.d. 1.5). The ratio 0.2 of the largest surviving Bjurböle meteorite to the projectile (terminal body) mass meets the catastrophic disruption threshold of 0.5 (Wickham-Eade, et al. 2018). Thus, the impact conditions such as impacting velocity (even though being subsonic with respect to material $v_p$ of 1030 m/s) and peak shock pressure were sufficient to catastrophically disrupt the projectile represented by the Bjurböle terminal body in a single event and the studied Bjurböle meteorites are representative of projectile catastrophic disruption.

There are three unique factors in our Bjurböle fragment study where current datasets are sparse or non-existent. First, the Bjurböle meteorite breakup represents an opportunity to study collision fragments in dm size range positioned in-between experimentally produced fragments (below ~cm) and asteroid boulders (above ~m). Second, Bjurböle is a chondritic impactor, being more relevant to planetary collisions when compared to basaltic, shale, plastic, or metallic impactors used in experiments. Third, the Bjurböle material is relatively weak and porous compared to other rocky materials (e.g., basalt, shale) used in most previous projectile break-up studies providing data on friable material breakup.

Nevertheless, we are aware that two factors may limit the applicability of our results. First, the estimated terminal impact velocity of 145-200 m/s is roughly 25-35 times lower than the typical 5 km/s collision velocity within the asteroid belt (Bottke et al. 1994). In our case this is advantageous as it is sufficient to achieve catastrophic mechanical breakup of the terminal body and at same time avoid internal pore collapse, impact melting, or other shock features associated with higher impact velocities(Guldemeister et al. 2022). Second, there is an uncertainty in the Bjurböle impactor original shape. However, (Durda, et al.

2015); Michikami, et al. (2016); (Wickham-Eade, et al. 2018) reported that the projectile or target shape have little effect on fragment shapes or sizes if any at all.

An interesting aspect of our results are the shapes and dimension ratios of Bjurböle fragments. The mean *b/a* and *c/a* values (0.84 and 0.67, respectively) are slightly higher compared to these reported from target catastrophic disruption experiments (~0.7 and ~0.5, respectively) for wide range of material compositions and porosities (Capaccioni, et al. 1984; Fujiwara, Kamimoto, & Tsukamoto 1978; Lange & Ahrens 1981; Michikami, et al. 2016)) meaning that our projectile fragments are closer to being equidimensional than typical target disruption fragments. This trend is also supported by a slightly larger mean apparent aspect ratio (0.77 with a s.d. of 0.11) compared to mean values of boulders observed on (65803) Didymos I Dimorphos (secondary), (25143) Itokawa, (162173) Ryugu, and (101955) Bennu asteroids (compare to asteroid range of 0.70-0.75 in Table 3 based on bounding box-derived data by Robin (2024)). The apparent aspect ratio of these boulders on the surface of asteroids has been associated to be formed by catastrophic disruption (Michikami, et al. 2019; Michikami, et al. 2010; Robin 2024).

Morphological parameters such as the circularity (~0.81 with s.d. of 0.05, large scale roughness indicator) is slightly smaller and solidity (~0.97 with s.d. of 0.02, small scale roughness indicator) is slightly higher compared to mean asteroid values revealing rather less concave shape of the Bjurböle fragments (Fig. 6, Table 3, asteroid range based on data by Robin (2024)). However, the mean circle ratio sphericity of Bjurböle fragments is in the same range as boulders on the surface of asteroids (Table 3), indicating a similar boulder elongation. The mean roundness of Bjurböle fragments indicates a rather angular medium (0.33 with a s.d of 0.04), which suggests that the fragments undergone limited alteration processes after the fragmentation at impact. The comparison with asteroid boulders is difficult, as the roundness value obtained for these boulders only provides a higher limit (< 0.56,(Robin 2024)) due to the lower image resolution of asteroid surfaces.

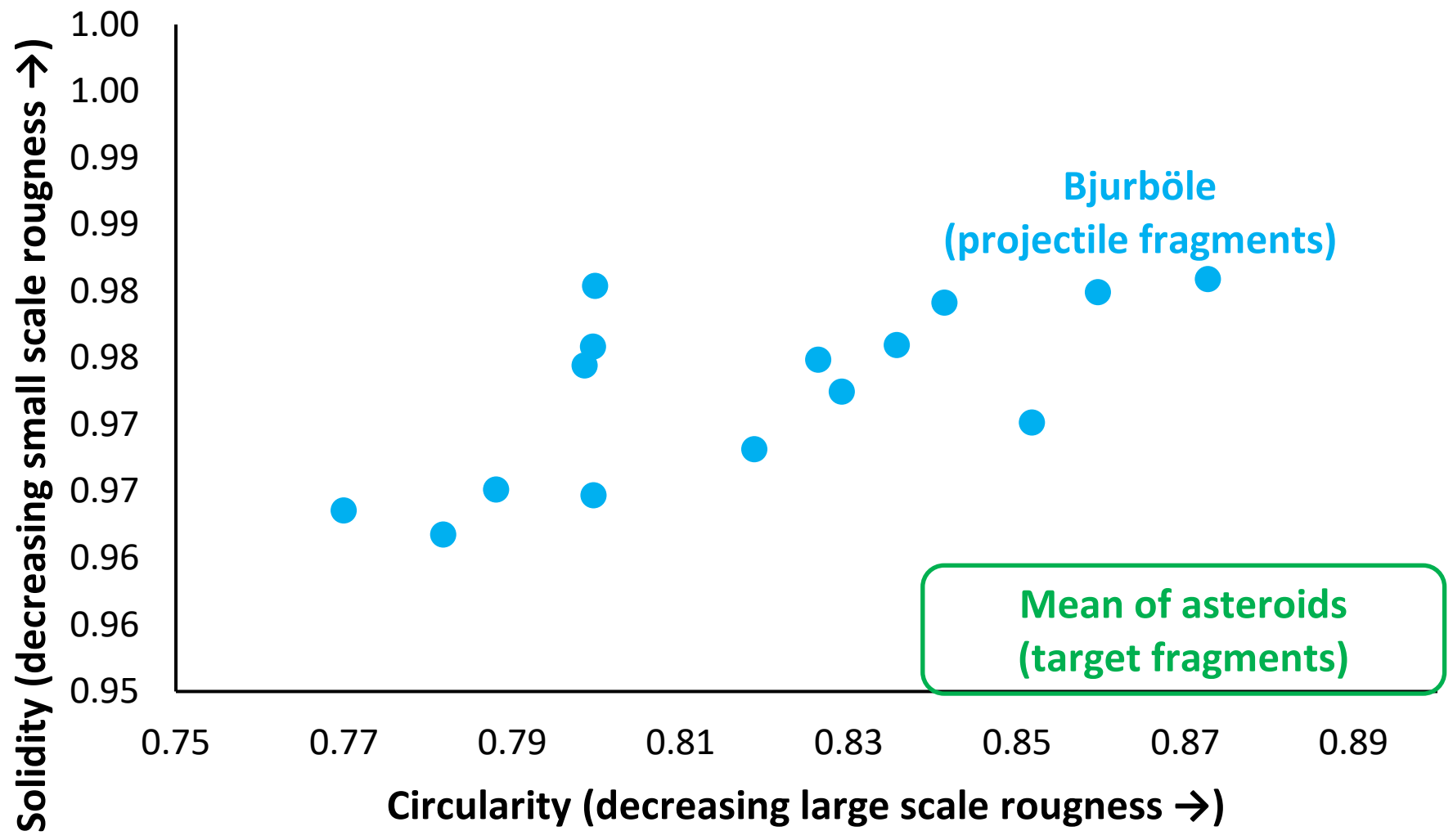

Fig. 6. Comparison of Bjurböle fragment compactness and solidity to these observed on asteroid boulders (Robin 2024).

Asteroid boulders are most likely predominantly composed of target fragments but may contain significant fraction of projectile fragments depending on the collision event circumstances. The fragment apparent aspect ratio derived from random 2D projections is a suitable proxy to similar asteroid boulder measure from 2D images. Indeed, the Bjurböle mean apparent aspect ratio value of 0.77 is slightly higher to asteroid Ryugu (~0.70), (433) Eros (~0.72), Itokawa (~0.62-0.74 depending on methodology), Bennu (~0.72), and Dimorphos (~0.60-0.70), boulder values reported in (Michikami & Hagermann 2021; Michikami, et al. 2019); Michikami, et al. (2010); (Pajola 2024) despite the asteroid boulders being one (Ryugu, Itokawa, Bennu, Dimorphos) or two (Eros) orders of magnitude larger than our Bjurböle meteorite fragments. The abovementioned asteroid values are in-between Bjurböle apparent aspect ratio value of 0.77 and mean true *c/a* ratio of 0.67 and often show bimodal apparent aspect ratio value distribution (Michikami, et al. 2010; Pajola 2024) similarly to our Bjurböle *c/a* histogram showing long asymmetric shoulder towards larger values (Fig. 4). This observation suggests that some boulders on these asteroids may be resting on their *a-c* plane. On the contrary, boulders on the asteroids with higher apparent aspect ratio values and unimodal distribution (similar to that of Bjurböle *b/a* values) may rest rather on their *a-b* plane. Asteroid (65803) Didymos (primary) show high boulder (< 16.5 m) mean apparent aspect ratio value of 0.86 (Pajola 2024) being even slightly higher than that of mean Bjurböle *b/a* ratio meaning that Didymos boulders are more equidimensional compared to Bjurböle or boulders of Dimorphos or other asteroids. This is an interesting observation given the generally accepted conclusion (Barnouin 2024) that Dimorphos is formed from mass shedding and gravitational re-accumulation of a material originating from Didymos, being itself a member of the Baptistina asteroid family, and may indicate a presence of a fragment sorting mechanism influencing Dimorphos formation (Pajola 2024).

We admit that the above-mentioned interpretations are speculative and require future detailed 3D study of asteroid boulders. We note that (Michikami & Hagermann 2021; Michikami, et al. 2019); Michikami, et al. (2010); (Pajola 2024) observe systematic boulder apparent aspect ratio increase with decreasing size and attribute this to re-orientation of smaller boulders on the surface due to impact shaking. The relatively high Bjurböle apparent aspect ratio values that we observe at decimeter scale highlight potentially important difference between collisional target and projectile (Bjurböle meteorite) disruption. This is supported by Bjurböle mean apparent aspect ratio value being good match to mean apparent aspect ratio values of sub-mm projectile fragments ($10^1$-$10^2$ μm) reported by Wickham-Eade, et al. (2018) for impact velocities below 1 km/s. The general trend of higher apparent aspect ratio is more pronounced for the weaker shale material compared to stronger basalt. Our results validate this trend up to the decimeter size range. This observation is further supported by Bjurböle relatively higher sphericity, and lower eccentricity. Michikami, et al. (2010) points out that boulders in the 0.1-5 m range identified in high-resolution Itokawa images (being similar in size to our fragments) appear less elongated with statistically higher apparent *b/a* ratios compared to the global average of larger boulders indicating that similar trend may exist with ordinary chondrite asteroid boulders.

Another effect affecting Bjurböle fragment morphology and shapes can be either presence of fusion crust or fragment erosion during penetration of the sea sediment or subsequent fragment cleaning. Bjurböle meteorites contain very little irregular fusion crust patches, mainly on the larger fragments (Table 1 and photos in on-line repository) as the break-up occurred during collision with sea ice rather than during hypervelocity atmospheric entry. Thus, atmospheric ablation cannot itself explain the observed differences. The fragment erosion may cause an observed change towards smoother and rounder appearance. However, its extent is unknown and impossible to quantify.

The Bjurböle meteorite cumulative "size" distribution (*a* axis) power index $\alpha$ of -2.7 is below the value of -2 which is considered to be indicative of prime catastrophic disruption with little subsequent fragmentation (Michikami & Hagermann 2021) and is comparable to experiments with basalt projectiles impacting at velocities below 1 km/s (Wickham-Eade, et al. 2018) or slightly steeper compared to the one reported in Avdellidou, et al. (2016) for impact of rocky projectiles onto ice targets. Our Bjurböle meteorites are naturally free of thermal cracking, fatigue, or other natural fragmentation or accumulation process besides accidental or intentional specimen breakup during handling consistent with shallow $\alpha$ value even at small sizes. Similarly, cumulative "mass" distribution $\alpha$ values around ~1 are reported for target catastrophic disruption in impact experiments (Davis & Ryan 1990; Durda, et al. 2015; Fujiwara, et al. 1977; Michikami, et al. 2016). The break in the power law fits at 130 mm or 1135 g as indicated by $x_{min}$ may be caused either by the incompleteness of our fragment set below this limit or by the change in impact regime .(Takagi, Mizutani, & Kawakami 1984).

From asteroid observations, a similar cumulative "size" distribution $\alpha$ value of −3.05 ± 0.14 was reported for global Itokawa (Michikami & Hagermann 2021), -2.62 ± 0.19 for the Itokawa "head" (Michikami & Hagermann 2021), -2.66 ± 0.05 for asteroid Ryugu (Michikami & Hagermann 2021), -2.9 ± 0.3 asteroid Bennu (DellaGiustina, et al. 2019), or -3.4 ± 1.2 for Dimorphos (Pajola 2024). The shallower Itokawa "head" $\alpha$ value is explained by fine material migration occurring on Itokawa towards accumulation areas on its "neck" and "body" (Michikami & Hagermann 2021), collisional size segregation (Shinbrot et al. 2017), or seismic segregation (Sánchez, Scheeres, & Quillen 2022).

Higher cumulative "size" distribution $\alpha$ values are associated with higher projectile impact velocities (Wickham-Eade, et al. 2018) or are indicative of other processes responsible for subsequent fragmentation or material sorting. Some asteroids indeed show higher values as −3.25 ± 0.14 for Eros (Michikami & Hagermann 2021) or -3.6 ± 0.7 for Didymos (Pajola 2024). The origin of detected large boulders on Eros is assumed to be related to a large impact creating Shoemaker crater (Michikami & Hagermann 2021). On the other hand, while there is evidence of thermal cracking on Bennu (DellaGiustina, et al. 2019) and it may also occur on Ryugu, the resulting fragments are likely below the 30-cm image resolution, and thus do not steepen the observed global power index.

The binary asteroid Didymos – Dimorphos is a peculiar case. Didymos cumulative "size" distribution $\alpha$ value of -3.6 ± 0.7, with a 22.8 m boulder size limit, is similar to Dimorphos $\alpha$ value of -3.4 ± 1.3 with 5 m boulder size limit (Pajola 2024). One possible explanation is that Didymos was formed from an impact-disrupted rubble pile asteroid and that Dimorphos inherited to a large extent its fragment size distribution from parent Didymos through shedding and reaccumulation (Barnouin 2024). However, the Dimorphos boulder size frequency distribution does follow rather a Weibull distribution than a power law at boulder sizes smaller than 5 m (Pajola 2024) which is indicative of some subsequent process affecting the ~m boulder population. Additional impact fragmentation or thermal cracking may be responsible for the destruction of boulders smaller than a few meters below current imagery resolution shallowing the distribution. There is evidence of thermal cracking from close-up observations of boulder morphologies of Dimorphos (Lucchetti 2024). However, with respect to young 0.03 - 13.3 Ma age of Dimorphos (Barnouin 2024) the thermal cracking simulations (Lucchetti 2024) require longer timescales to destroy small boulders and thus, may be not primarily responsible for observed depletion of ~m-sized boulders. Our results indicate that the shallow $\alpha$ of -2.9 at sub-meter sizes is also compatible with a single impact disruption at collision velocities below ~1 km/s. Unfortunately, no similar high-resolution observations are available on Didymos so one cannot detect or discard similar depletion in ~ m-sized boulders on the primary asteroid.

## 5. Conclusions

The relatively slow subsonic (~1/5 of material $v_p$) impact velocity of Bjurböle projectile was sufficient to cause its catastrophic disruption in a single impact event. This was followed with little additional fragmentation as indicated by the 0.2 mass ratio of the largest surviving Bjurböle meteorite to the projectile (Bjurböle terminal body), by a moderate cumulative size distribution power-law index $\alpha$ of -2.7, and cumulative mass distribution $\alpha$ of -1.0. The Bjurböle meteorites representing the projectile fragments are homogeneous from cm (10 g) up to dm (80 kg) size range as revealed through the bulk density measurements. They are rather equidimensional compared to fragments typically originating from target material as observed in impact experiments or on asteroid surfaces (with asteroid boulder origin being assumed to be predominantly of the fragmented target material). When compared to the boulders observed on visited asteroids, the Bjurböle projectile fragments are also characterized with a slightly lower roughness at large scale and larger roughness at small scale. These differences can be attributed either to different fragment source (projectile vs. target), to high porosity and low strength of Bjurböle material, or to lower subsonic impact velocity compared to typical impact velocities within the asteroid belt.

Our results imply that efficient catastrophic disruption of projectiles can be achieved even at sub-km/s velocities, especially with weak materials. Projectile fragments are rather small compared to the original projectile size and may be well mixed within target fragment population. On the contrary, projectile fragments tend to be in certain cases (especially for weak projectiles and lower collision velocities) more spherical with smoother large scale morphology compared to target fragments. No additional significant differences were observed between our decimeter-sized fragment properties and experimentally produced smaller fragments or naturally occurring larger asteroid boulders.

## 6. Acknowledgements

Authors would like to thank following individuals (listed in alphabetic order) for providing information on or access to meteorite samples in worldwide collections: Dmitry Badjukov, Solveig Bergholm, Natalia Bezaeva, Chloé Brillartz, Rainer Bartoschewitz, Neil Bowles, Kelsey Falquero, Henrik Friis, Matthieu Gounelle, Victor Grokhovsky, James Holstein, Svetlana Janson, Juha Jämbäck, Jarkko Kettunen, Kitty Killgore, Andrei Kosterov, Nikolai Kruglikov, Jörgen Langhof, Jussi Leveinen, Jarmo Moilanen, Emma Nicholls, David Rose, Mikhail Shabalov, Reidar Tronnes, Grigoriy Yakovlev, Luminita Zaharia, and Jutta Zipfel. Information on location of Bjurböle material in worldwide collections was gathered from LUOMUS and MetBase databases. This work was supported by the DART mission, NASA Contract 80MSFC20D0004, Academy of Finland project no. 335595, and institutional support RVO 67985831 of the Institute of Geology of the Czech Academy of Sciences. This work was supported by the Italian Space Agency (ASI) within the LICIACube project (ASI-INAF agreement n. 2019-31-HH.0) and HERA project (ASI-INAF agreement n. 2022-8-HH.0). A.D and N.M. acknowledge funding support from the European Commission's Horizon 2020 research and innovation programme under grant agreement No 870377 (NEO-MAPP project), and the Centre National d'Etudes Spatiales (CNES), focused on the Hera space mission. A.D. acknowledges PhD funding from University of Toulouse III.

## 7. Author contributions

TK led the work, participated in the meteorite measurements, and did the majority of data interpretation. AS participated on the meteorite measurements and contributed with data interpretation (density and porosity). AL contributed with gathering and analysis of Bjurböle historic records. AD and NM contributed with fragment shape analysis. MP and AL contributed with the fragment SFD and MFD and derived the best fit, contributed to part of the manuscript and interpretation. RL contributed with peak shock pressure estimate. NLC, SDR PS, OSB, and ASR contributed to data interpretation.

## 8. Data availability

Meteorite measurement datasets, 3D shape models, together with metadata are available in the Zenodo repository at https://www.doi.org/10.5281/zenodo.10062980.

## Supporting data

Supplementary Table S1: Bulk and grain density and porosity together with global mean of all laser-scanned Bjurböle fragments including additional smaller meteorites. Additionally are listed values of Helsinki A3747 meteorite plaster model, Helsinki A3747 real meteorite with watertight approximation of its invisible bottom part, and Chelyabinsk meteorites from (Kohout, et al. 2014; Kohout, et al. 2017) presented for comparison in Fig. 3. Note that in case of Bjurböle the grain density mean is calculated from samples Helsinki small 1-4 only and the mean value is used for other Bjurböle samples (indicated by italics) to estimate porosity. Similarly, the mean grain density of Chelyabinsk small meteorites is used to estimate porosity of the three large meteorites.

| Meteorite | Mass (g) | Bulk Volume ($cm^3$) | Bulk Density ($g/cm^3$) | Grain Volume ($cm^3$) | Grain Density ($g/cm^3$) | Porosity (%) |
|---|---|---|---|---|---|---|
| Helsinki A3747 (4) | 80200 | 29172.40 | 2.75 | | *3.53* | 22.2 |
| Stockholm 990433 | 20900 | 7302.16 | 2.86 | | *3.53* | 19.0 |
| Helsinki D6147 (AL79) | 13480 | 4860.55 | 2.77 | | *3.53* | 21.5 |
| Helsinki D6149 (AL86 / 33 5413) | 5450 | 1877.31 | 2.90 | | *3.53* | 17.9 |
| Helsinki D6150 (TK1) | 5300 | 1862.90 | 2.85 | | *3.53* | 19.5 |
| Helsinki D6164 (AL80 / 5318) | 5150 | 1818.94 | 2.83 | | *3.53* | 19.9 |
| Helsinki D6152 (AL80) | 4500 | 1585.59 | 2.84 | | *3.53* | 19.7 |
| Helsinki D6148 (AL81) | 4350 | 1518.40 | 2.86 | | *3.53* | 18.9 |
| Helsinki D6163 (3130 / 26) | 3113.7 | 1082.62 | 2.88 | | *3.53* | 18.6 |
| Kettuen RH80 (Haag 107) without corner | 2977.6 | 1009.85 | 2.95 | | *3.53* | 16.6 |
| Helsinki D6151 (AL2 27) | 1879 | 665.49 | 2.82 | | *3.53* | 20.1 |
| Helsinki D6146 (AL68) | 1809.4 | 638.79 | 2.83 | | *3.53* | 19.8 |
| Helsinki D6144 (AL32 / 143) | 1704.8 | 589.85 | 2.89 | | *3.53* | 18.2 |
| Porvoo 9184 | 1432.8 | 503.64 | 2.84 | | *3.53* | 19.5 |
| Helsinki D6145 (AL67 / 152) | 1358.6 | 474.43 | 2.86 | | *3.53* | 19.0 |
| Aalto 4606 | 1364.7 | 468.56 | 2.91 | | *3.53* | 17.6 |
| Helsinki D6143 (AL31 / 144) | 1237.9 | 441.98 | 2.80 | | *3.53* | 20.7 |
| Stockholm 540022 | 1176.7 | 404.84 | 2.91 | | *3.53* | 17.8 |
| Porvoo 88-38 | 1069.6 | 377.45 | 2.83 | | *3.53* | 19.8 |
| Helsinki 1002 | 856.5 | 300.77 | 2.85 | | *3.53* | 19.4 |
| Helsinki AL63 | 754.4 | 263.77 | 2.86 | | *3.53* | 19.1 |
| Copenhagen 947 | 465.6 | 159.84 | 2.91 | | *3.53* | 17.6 |
| Helsinki AL34 | 151.75 | 53.59 | 2.83 | | *3.53* | 19.9 |
| Helsinki small 4 | 21.62 | 7.83 | 2.76 | 6.13 | 3.53 | 21.7 |
| Helsinki small 3 | 20.89 | 7.51 | 2.78 | 5.92 | 3.53 | 21.1 |
| Helsinki small 2 | 20.56 | 7.41 | 2.78 | 5.8 | 3.54 | 21.7 |
| Helsinki small 1 | 17.27 | 6.26 | 2.76 | 4.88 | 3.54 | 22.0 |
| Helsinki SC57 | 12.252 | 4.28 | 2.87 | | *3.53* | 18.9 |
| Mean | | | 2.84 | | 3.53 | 19.6 |
| s. d. | | | 0.05 | | 0.01 | 1.5 |
| | | | | | | |
| Helsinki A3747 (4) plaster model | 80200 | 29376.51 | 2.73 | | 3.53 | 22.7 |
| Helsinki A3747 (4) real watertight approximation | 80200 | 28968.30 | 2.77 | | 3.53 | 21.7 |
| | | | | | | |
| Chelyabinsk small average | | | 3.29 | | 3.47 | 6.0 |
| Chelyabinsk big | 502000 | 156122.2 | 3.22 | | 3.51 | 6.6 |
| Chelyabinsk S | 22750 | 6989.1 | 3.26 | | 3.51 | 6.1 |
| Chelyabinsk V | 5130 | 1581.61 | 3.24 | | 3.51 | 7.2 |